\documentclass[%
twocolumn,
superscriptaddress,
 amsmath,amssymb,
 aps,
prstab,
nofootinbib
]{revtex4-2}
\usepackage{graphicx}% Include figure files
\usepackage{dcolumn}% Align table columns on decimal point
\usepackage{bm}% bold math
\usepackage[USenglish]{babel}
\usepackage[separate-uncertainty]{siunitx}% Nice micron signs and units
\usepackage{multirow}
\usepackage{textcomp}
\usepackage{hhline}
\usepackage{tabularx}
\usepackage[flushleft]{threeparttable}
\usepackage[table]{xcolor}
\newcommand{\murm}{\hbox{\textmu}}

\DeclareUnicodeCharacter{308}{}
\usepackage{xcolor}
\definecolor{prlblue}{rgb}{0.176, 0.152, 0.57}
\usepackage[colorlinks=true, linkcolor=black, citecolor=prlblue, filecolor=black, urlcolor=prlblue]{hyperref}
\begin{document}

\title{A hybrid, asymmetric, linear Higgs factory \\ based on plasma-wakefield and radio-frequency acceleration}

\author{B.~Foster}
\email{brian.foster@physics.ox.ac.uk}
\affiliation{John Adams Institute for Accelerator Science at University of Oxford, Oxford, UK}
\affiliation{Deutsches Elektronen-Synchrotron DESY, Hamburg, Germany}
\author{R.~D'Arcy}
\affiliation{John Adams Institute for Accelerator Science at University of Oxford, Oxford, UK}
\affiliation{Deutsches Elektronen-Synchrotron DESY, Hamburg, Germany}
\author{C.~A.~Lindstr{\o}m}
\affiliation{Department of Physics, University of Oslo, Oslo, Norway}

\date{\today}

\begin{abstract}
The construction of an electron--positron collider ``Higgs factory" has been stalled for a decade, not because of feasibility but because of the cost of conventional radio-frequency (RF) acceleration. Plasma-wakefield acceleration promises to alleviate this problem via significant cost reduction based on its orders-of-magnitude higher accelerating gradients. However, plasma-based acceleration of positrons is much more difficult than for electrons. We propose a collider scheme that avoids positron acceleration in plasma, using a mixture of beam-driven plasma-wakefield acceleration to high energy for the electrons and conventional RF acceleration to low energy for the positrons. We emphasise the benefits of asymmetric energies, asymmetric bunch charges and asymmetric transverse  emittances. The implications for luminosity and experimentation at such an asymmetric facility are explored and found to be comparable to conventional facilities; the cost is found to be much lower. Some of the areas in which R\&D is necessary to make HALHF a reality are highlighted,including estimates for the improvement required in key technologies. These range from a factor of 10 to a factor of 1000. 

\end{abstract}


\maketitle

\section{introduction}
There is a consensus in the world-wide particle-physics community~\cite{CERN_Strategy_2020, Snowmass_2021} that the next energy-frontier particle collider should be an electron--positron ``Higgs factory", which would greatly extend our understanding of the most mysterious, and newest, of the elementary particles that make up what is known as the Standard Model of particle physics. 
Such a Higgs factory would produce copious amounts of Higgs bosons via the reaction 
\begin{equation*}
    e^+ e^- \rightarrow H Z,
\end{equation*}
which requires a center-of-mass (c.m.) energy in excess of 216~GeV. 
Although the LHC both discovered the Higgs boson and has made impressive determinations of many of its properties, which will also improve with future LHC running, it is widely accepted that an electron--positron Higgs factory, with its exceptionally clean experimental conditions, will greatly extend our knowledge beyond what even the full running envisaged for the High-Luminosity LHC can achieve~\cite{Snowmass_Higgs_Physics}. 

There is however no consensus on the best technology for such a Higgs factory. In brief,  
the relatively mature proposals, the International Linear Collider (ILC)~\cite{ILC_TDR_2013}, using superconducting niobium radio-frequency (RF) cavity structures and the Compact Linear Collider (CLIC)~\cite{CLIC_CDR_2012}, using normal-conducting copper accelerating structures, are expensive.
Circular colliders use mature technology but have to be very large to give reasonable synchrotron-radiation losses, so that the Circular Electron--Positron Collider (CEPC)~\cite{CEPC_CDR} is similar in capital construction cost to the linear colliders, while the Future Circular Collider (FCC--ee)~\cite{FCCee} is significantly more expensive.
A technology that could promise a Higgs factory at a greatly reduced cost would therefore be truly disruptive.

In recent years, the field of plasma-wakefield accelerators (PWFA) \cite{Veksler_HEACC_1956,Tajima_PRL_1979,Chen_PRL_1985,Ruth_PA_1985} has seen enormous advances. 
The demonstration of stable operation for many hours \cite{MaierPRX2020} and that MHz repetition rates are in principle allowed by basic plasma properties \cite{DArcyNature2022} open up the possibility of user-oriented devices coming into operation within the next decade, e.g. a plasma-driven free-electron laser \cite{Wang_Nature_2021,Pompili_Nature_2022} or a plasma injector to a storage ring \cite{Antipov_PRAB_2021}.

However, despite discussion going on for four decades \cite{Channell_AIP_1982,Lawson_Nature_1984}, as yet no PWFA device has contributed to particle-physics applications. 
There are several reasons for this. Energy-frontier devices such as electron--positron colliders probe point-like processes whose annihilation cross-sections fall like $1/s$, where $\sqrt{s}$ is the c.m. energy. 
To produce interesting numbers of events, the luminosity or, equivalently, beam power, must increase with energy to compensate. 
Even the most powerful current PWFAs, such as at FACET-II \cite{Yakimenko_PRAB_2019} and FLASHForward \cite{DArcy_RSTA_2019}, are still limited to beam powers far below those required for e.g.~the ILC \cite{ILC_TDR_2013}. 
This is even more true for laser-driven plasma-wakefield acceleration (LWFA), where the power efficiency of appropriate lasers is still many orders of magnitude below particle-physics requirements \cite{Wim_1,Wim_2,Wim_3,Wim_4,Hooker_JPhysB_2014}. 
For this reason, we confine our remarks below to beam-driven (PWFA) devices. 

Another problem, by no means the least difficult, is the acceleration of positrons in a PWFA device \cite{Lee_PRE_2001}. 
Almost all experimental work to date has been carried out with electron acceleration. Pioneering work at the FFTB and FACET facilities at SLAC \cite{Blue_PRL_2003,Hogan_PRL_2003,Muggli_PRL_2008,Corde_Nature_2015,Gessner_NatCommun_2016,Doche_SciRep_2017,Lindstrom_PRL_2018} has demonstrated positron acceleration, but reproducibility, stability, efficiency and beam parameters appropriate for particle-physics applications are not currently in sight. 
There are many proposals for possible positron-acceleration mechanisms \cite{Vieira_PRL_2014,Diederichs_PRAB_2019,Hue_PRR_2021,Silva_PRL_2021,Zhou_PRL_2021}. However, experimental demonstration, hopefully possible in the future at FACET-II, seems likely to be many years away. Application in a user-oriented facility is even further in the future.

The above situation is particularly unfortunate in that, currently, progress towards a collider based on conventional radio-frequency cavity technology, such as ILC or the Compact Linear Collider (CLIC) \cite{CLIC_CDR_2012} is very slow.
This is not due to technology limitations~--~with a few minor exceptions, the ILC Technical Design Report represents a “shovel-ready” project~--~but more the political complications directly related to the perceived high cost of these facilities. Even the most recent, and still to mature, technology proposed, 
the Cool Copper Collider~\cite{CCC}, 
is estimated by the Snowmass Collider Implementation Task force to have a similar cost to ILC and CLIC~\cite{Snowmass_Implementation}.

It is precisely in cost reduction that PWFA could play a decisive role. 
The very high gradients attainable promise a significant reduction in the length of the main linear accelerator and hence a concomitant reduction in cost~\cite{Adli_PWFA_LC}. 
While there is realistic hope that a particle-physics-oriented PWFA electron accelerator could soon be  a reality, no analogous situation currently pertains for a plasma-based positron accelerator.

This paper addresses the current incompatibility of plasma accelerators and electron--positron colliders. 
Rather than wait an unspecified but clearly long time for positron PWFA acceleration to be solved, we propose a hybrid, asymmetric, linear Higgs Factory (HALHF), in which electrons are accelerated to higher energy in PWFAs and positrons are accelerated to lower energy in conventional RF cavities. 
In order to maximise the potential cost reduction for such a device compared to ILC or CLIC, the positron energy must be considerably lower than in either of those proposals. 
To compensate and obtain a c.m. energy sufficient to produce Higgs bosons, the electron energy must be appropriately higher, taking advantage of the high accelerating gradients in PWFA. 
This means that the c.m. of the electron--positron annihilations will be boosted in the direction of motion of the electron beam. 

In the following, we first discuss the criteria on which the choice of suitable beam energies and bunch charges for the HALHF can be made, and the effect of these choices on energy efficiency and luminosity per bunch crossing. 
The effect of beam--beam limitations on the latter are investigated using GUINEA-PIG~\cite{Schulte_Thesis_1996}. 
An approximate footprint and schematic layout are shown. 
The characteristics of the electron beam from a PWFA-based accelerator of the required energy, including the number of PWFA stages required and the overall bunch-train pattern, are outlined. 
This is followed by  details on some aspects of experimentation at HALHF that differ from those at a symmetric collider. 
We then outline a capital cost estimate and estimate the running costs, based predominantly on scaling from other proposals. 
The penultimate section details staging and upgrade options. In the conclusions, we also remark on the substantial improvements in performance that are necessary in the proposed R\&D programme.

\section{Beam-energy specification}
\label{sec:BE-spec}

The couplings of the Standard Model are such that the lowest c.m. energy necessary for Higgs-boson production is the Higgs--$Z$ threshold at 216.4~GeV. 
In order to scan beyond threshold and give a safety margin, a minimum running energy of 250~GeV is usually chosen for Higgs factories. 
We will follow this for the energy of HALHF. 

Simple relativistic kinematics determines the choice of beam energies once a number of other factors are taken into consideration. 
The most important is to minimise the cost of the facility. 
Since conventional RF linac technology is expensive, this is principally achieved by minimising the cost of the positron linac. 
The lower limit of the positron energy is chosen to optimise efficiency and facility footprint while giving a c.m. energy boost, $\gamma$,
similar to that produced at the HERA electron/positron--proton collider, where a very successful programme of experimentation was carried out~\cite{HERA}. 
This gives confidence that a similarly successful detector could be designed for HALHF (see Section~\ref{sec:Exp}). 

The relativistic relations
\begin{equation} 
    E_e E_p = s/4 
\end{equation}
and
\begin{equation}
    \label{eq:totalenergy}
    E_e + E_p = \gamma \sqrt{s},
\end{equation}
where $E_e$ and $E_p$ are the electron and positron energies, respectively, govern the kinematics. 
These two equations link three variables; fixing one therefore determines the other two. For a given choice of positron and c.m. energy, the boost becomes
\begin{equation}
    \gamma = \frac{1}{2}\left(\frac{2E_p}{\sqrt{s}} + \frac{\sqrt{s}}{2E_p}\right).
\end{equation}
As $\gamma$ increases, products of the annihilation are boosted into a narrowing cone around the more energetic lepton direction, so that it eventually becomes experimentally problematic to disentangle the final states.
Based on a crude optimisation of the overall facility length, we choose a positron energy of 31.3~GeV, four times lower than the symmetric case. This leads to an electron energy four times higher (500~GeV) and a boost of $\gamma = 2.13$, smaller than the value of 3 at HERA and therefore more favourable for experimentation. 
We use these values in the following to give indicative properties and cost for HALHF.

\begin{table*}[t]
    \setlength\tabcolsep{0pt}
    \centering
    \begin{tabularx}{\textwidth}{|@{}>{\centering}X@{}|@{}>{\centering}X@{}|@{}>{\centering}X@{}|@{}>{\centering}X@{}|@{}>{\centering}X@{}|@{}>{\centering}X@{}|@{}>{\centering}X@{}|@{}>{\centering}X@{}|@{}>{\centering}X@{}|c|} \hline
        \rowcolor{gray!10} $E$ (GeV) & $\sigma_z$ (\murm m) & $N$ ($10^{10}$) & $\epsilon_{nx}$ ($\murm$m) & $\epsilon_{ny}$ (nm) & $\beta_x$ (mm) & $\beta_y$ (mm) & $\mathcal{L}$ ($\murm\rm{b}^{-1}$) & $\mathcal{L}_{0.01}$ ($\murm\rm{b}^{-1}$) & \hspace{3pt}$P/P_0$\hspace{3pt}\hspace{0pt} \\ \hline \hline
        125 / 125   &  300 / 300  &  2 / 2  &  10 / 10  &  35 / 35   &  13 / 13    &  0.41 / 0.41  &  1.12  &  0.92  &  1     \\ \hline
        31.3 / 500  &  300 / 300  &  2 / 2  &  10 / 10  &  35 / 35   &  3.3 / 52   &  0.10 / 1.6   &  0.93  &  0.71  &  2.13  \\ \hline
        31.3 / 500  &  75 / 75    &  2 / 2  &  10 / 10  &  35 / 35   &  3.3 / 52   &  0.10 / 1.6   &  1.04  &  0.71  &  2.13  \\ \hline
        31.3 / 500  &  75 / 75    &  4 / 1  &  10 / 10  &  35 / 35   &  3.3 / 52   &  0.10 / 1.6   &  1.04  &  0.60  &  1.25  \\ \hline
        31.3 / 500  &  75 / 75    &  4 / 1  &  10 / 40  &  35 / 140  &  3.3 / 13   &  0.10 / 0.41  &  1.01  &  0.58  &  1.25  \\ \hline
        31.3 / 500  &  75 / 75    &  4 / 1  &  10 / 80  &  35 / 280  &  3.3 / 6.5  &  0.10 / 0.20  &  0.94  &  0.54 &  1.25  \\ \hline
        31.3 / 500  &  75 / 75    &  4 / 1  &  10 / 160 &  35 / 560  &  3.3 / 3.3  &  0.10 / 0.10  &  0.81  &  0.46 &  1.25  \\ \hline
        \hline
        45.6 / 45.6 &  109 / 109  &  2 / 2       &  10 / 10  &  35 / 35   &  4.7 / 4.7 &  0.15 / 0.15  &  1.12  &  0.93  & 1 \\ \hline
        31.3 / 66.5 &  75 / 75    & 2.9 / 1.4  &  10 / 21  &  35 / 75  &  3.3 / 3.3 &  0.10 / 0.10   &  1.06  &  0.78  & 1.07 \\ \hline
        11.4 / 182  &  27 / 27    &  4 / 1       &  10 / 160  &  35 / 560 &  1.2 / 1.2 &  0.04 / 0.04   &  0.81  &  0.46  &  1.25 \\ \hline
    \end{tabularx}
    \caption{GUINEA-PIG simulations showing the luminosity per bunch crossing of both symmetric and asymmetric collisions. The first number in each pair refers to the positron bunch, the second to the electron bunch. Tabulated are, from left to right, beam energies, bunch lengths, number of particles per bunch, normalised emittances in the horizontal and vertical planes, interaction-point beta functions in the horizontal and vertical planes, the calculated full luminosity and that with energy within
    1\% of the nominal peak in inverse microbarns, and the relative power increase required compared to symmetric collisions. Numbers in the upper table represent $HZ$ operation, whereas the lower table represents $Z$ operation. The first row in each section of the table represents ILC-like parameters. Simulations include a vertical waist shift (equal to the bunch length), but assume zero transverse offsets and crossing angle. }
    \label{tab:1}
\end{table*}

\section{Effect on energy efficiency}
The use of asymmetric beam energies leads to a reduction in the energy efficiency.
The total energy required to collide two bunches is given by the sum of their energies, $N_eE_e + N_pE_p$, where $N_e$ and $N_p$ are the numbers of electrons and positrons per bunch, respectively.
Assuming equal charges in the electron and positron bunches ($N = N_e = N_p$), symmetric collisions require a total energy of
$2N \sqrt{s}$, while asymmetric collisions, as seen from Eq.~\ref{eq:totalenergy}, require $2\gamma N \sqrt{s}$.
Thus the asymmetry chosen here implies an increase in energy usage by a factor $\gamma=2.13$ compared to the symmetric case.

However, in the energy-asymmetric case, the loss in energy efficiency can be mitigated, and even fully cancelled, by also introducing a charge asymmetry~--~i.e., reducing the number of particles in the high-energy electron beam and increasing the number in the low-energy positron beam.
To maintain luminosity, the product of the total bunch charges ($N^2 = N_e N_p$) must remain constant.
With this constraint, the relative increase in power is given by
\begin{equation}
    \frac{P}{P_0} = \frac{N_e E_e + N_p E_p}{N\sqrt{s}},
\end{equation}
where $P_0$ and $P$ represent the power required in symmetric and asymmetric collisions, respectively.  
The same energy efficiency can be achieved in asymmetric as in symmetric collisions by scaling the charge of each bunch inversely proportional to its change in energy: $N_e/N = N/N_p = 2 E_p/\sqrt{s}$.
In our case, this corresponds to a factor 4.
This may in practice be difficult to reach due to constraints on positron production.
Using instead a factor 2 (double the positrons, half the electrons) nearly doubles the energy efficiency compared to equal-bunch-charge collisions, making the proposed asymmetric collider only 25\% less energy efficient than an energy-symmetric machine. 
However, there is a small drop in luminosity for such a configuration, as discussed in the next section.

In addition to increasing the asymmetry in bunch charge, there may be scope in plasma-accelerator technology for improving the energy efficiency\footnote{It has been suggested that the energy efficiency in a plasma accelerator may be limited in practice by transverse instabilities~\cite{LebedevPRAB2017}~--~an issue subject to much recent research and many proposed solutions~\cite{LehePRL2017,Mehrling2018}, but no consensus has yet been reached regarding the maximum achievable efficiency.}, as beam-driven plasma accelerators can, in principle, be more energy efficient than conventional RF technology \cite{Tzoufras_beamloading,Adli2013}.

%--- FIGURE 1: setup figure ---%
\begin{figure*}[t]
	\centering\includegraphics[width=\textwidth]{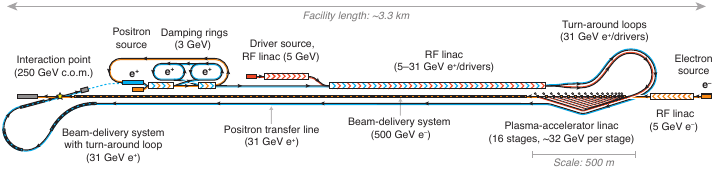}
	\caption{Schematic layout of the hybrid asymmetric linear Higgs factory. Particle sources provide electrons (orange), positrons (blue) and electron drivers (red) for acceleration. Electrons are accelerated to 5~GeV and diverted via a return loop back to produce the positrons in the positron-source complex. The positrons are then captured, accelerated to 3~GeV and injected into a pre-damping ring. A second damping ring produces low-emittance bunches that are accelerated in the remaining RF linac up to 31.3~GeV, before being turned around and sent down a transfer line (the kink in the positron transfer line after the first turn-around is not required but drawn for clarity~--~the line can pass either below or above the undulating delay chicanes). The positrons enter the beam-delivery system, which is combined with another turn-around, and then enter the final focus to collisions. Trains of electron drivers are accelerated, first to 5~GeV in a dedicated RF linac, then accelerated after the positron bunch to 31.3~GeV. After transfer to a turn-around loop, the drive beams are separated and injected into the appropriate PWFA stages, sixteen in total, where each sequentially accelerates a electron bunch from a photocathode injector up to 500~GeV. The spent electron drivers are discarded into separate beam dumps. The accelerated electron bunch enters the high-energy beam-delivery system and collides with the positron bunch. The spent colliding beams enter beam dumps located after the interaction point. The dashed line represents an option to use the spent positrons for positron production. The approximate length of the facility is 3.3~km. (Note that a detailed implementation of the HALHF concept is shown in Fig.~\ref{fig:3_bunchtrainpattern} and discussed thereafter).}
    \label{fig:1_setup}
\end{figure*}
 
\section{Effect on luminosity per bunch crossing}
\label{sec:lumiperX}
Asymmetric beam energies do not affect the geometric luminosity as long as beam sizes remain constant.
Nevertheless, two important differences exist between a symmetric and an asymmetric collision for a given beam size: the ``hour-glass" effect \cite{Furman_SLAC} and the beam--beam effect \cite{Schulte_BB_2017}.

To make the appropriate comparison, beam sizes can be kept constant by scaling the interaction-point (IP) beta functions by the square root of the energy.
Since, compared to the symmetric case, the low-energy positron bunch will have a larger geometric emittance, which scales inversely with beam energy,  this must be compensated by a smaller beta function.
Fortunately, there is scope to decrease IP beta functions below what has been proposed for ILC ($13 \times0.41$~mm), to close to a proposal for CLIC ($8 \times 0.1$~mm)~\cite{Garcia_CLIC}.

The hour-glass effect therefore dictates that the positron-bunch length must be reduced by a similar factor. The inverse is true for the high-energy electron bunch: the interaction-point beta functions can be increased compared to the symmetric case, which could reduce the complexity and cost of the beam-delivery system (see Section~\ref{sec:bds}).
However, a far more intriguing option is to increase the emittance.
If additionally the beta functions are scaled down (similar to that of the positrons), the emittances can be doubly scaled up~--~if the electron energy is four times higher than the symmetric case, the normalised emittance can be as much as sixteen times higher in both planes.
This observation is of great importance since the electrons are accelerated using PWFAs, where it may be problematic to maintain nm-scale normalised emittances.
The introduction of an energy asymmetry therefore increases the operational tolerance for emittance growth in the PWFA arm.

The beam--beam focusing effect that leads to the well-known luminosity enhancement \cite{Schulte_BB_2017} is expected to be reduced for the high-energy beam and increased for the low-energy beam for a given bunch length.
However, the overall effect on the luminosity is non-trivial and must be simulated.
Using GUINEA-PIG \cite{Schulte_Thesis_1996}, we compared the luminosity for a situation with parameters similar to ILC~\cite{ILC_TDR_Accelerator} with that of an asymmetric collider as discussed above.
Table~\ref{tab:1} shows that the luminosity assuming ILC bunch lengths (300~{\murm}m rms) decreases by approximately 17\% (23\% for the luminosity within 1\% of the peak of the spectrum).
Reducing the bunch length by a factor of 4 to compensate for the hour-glass effect gives a smaller luminosity reduction of only approximately 7\% (23\%).
Introducing a charge asymmetry for energy efficiency changes this to a 7\% (35\%) luminosity drop.
Furthermore, using various degrees of asymmetric emittances, ranging from a factor 4 to 16, decreases the luminosity per bunch crossing from the ILC TDR values by between 10\% (37\%) and 28\% (50\%). In order to benefit from the less stringent tolerances in the PWFA arm, we assume the parameters given in the final row of the top half of Table~\ref{tab:1} (summarised in Table~\ref{tab:2}).

Also shown in Table~\ref{tab:1} is a comparison of the HALHF and ILC luminosities per bunch crossing at the peak of the $Z$ resonance for two situations: firstly where the positron-beam energy is maintained at 31.3~GeV and the electron energy set accordingly, which gives a boost of 1.07; secondly where the boost is maintained at 2.13. 
In the former case, the HALHF luminosity per bunch crossing is reduced by 5\% (16\%), while in the latter it is reduced by approximately 28\% (50\%), compared to ILC.

\section{Schematic layout of the collider}
\label{sec:Setup}
This section gives a broad-brush description of the main components of HALHF, as shown in Fig.~\ref{fig:1_setup}.
In order to make comparisons easier, we re-use as much as possible of existing designs, mostly from ILC and CLIC.

The HALHF structure contains two main accelerators~--~one consisting of a conventional linac for electron drivers and positrons, and the other a plasma-based linac for the colliding electrons~--~and three particle sources: one for electrons to hit the positron target and produce the positrons, one for the electron drivers, and one for the colliding electrons. 
The remaining components of the facility include positron damping rings, beam-delivery systems, and transfer lines. 
These aspects are discussed in more detail below.
A summary of the HALHF parameters used is shown in Table~\ref{tab:2}.

\begin{table}[!htbp]
\begin{threeparttable}

    \begin{tabular}{p{0.51\linewidth}>{\centering}p{0.13\linewidth}>{\centering}p{0.15\linewidth}>{\centering\arraybackslash}p{0.15\linewidth}}
    \vspace{0.5cm}\\
    \textit{Machine parameters} & \textit{Unit} & 
    \multicolumn{2}{c}{\textit{Value}}\\[3pt]
    \hline
    & & \\[-15pt]
    Center-of-mass energy & GeV & \multicolumn{2}{c}{250} \\
    Center-of-mass boost &  & \multicolumn{2}{c}{2.13} \\[-7pt]
    Bunches per train & & \multicolumn{2}{c}{100} \\[-7pt]
    Train repetition rate & Hz & \multicolumn{2}{c}{100} \\[1pt]
    Average collision rate & kHz & \multicolumn{2}{c}{10} \\
    Luminosity & cm$^{-2}$~s$^{-1}$ & \multicolumn{2}{c}{$0.81\times10^{34}$} \\
    Luminosity fraction in top 1\% & & \multicolumn{2}{c}{57\%} \\[-7pt]
    Estimated total power usage & MW & \multicolumn{2}{c}{100} \\[3pt]
    \hline
     & & & \\[-15pt]
    \textit{Colliding-beam parameters} & & $e^{-}$ & $e^{+}$ \\[-6pt] 
    \hline 
    & & \\[-15pt]
    Beam energy & GeV & 500 & 31.25 \\
    Bunch population & $10^{10}$ & 1 & 4 \\
    Bunch length in linacs (rms) & $\mu$m & 18 & 75 \\ %Bunch length in linac (rms) & $\mu$m & 9 & 75 \\
    Bunch length at IP (rms) & $\mu$m & \multicolumn{2}{c}{75} \\
    Energy spread (rms) & \% & \multicolumn{2}{c}{0.15} \\
    Horizontal emittance (norm.) & $\mu$m & 160 & 10 \\
    Vertical emittance (norm.) & $\mu$m & 0.56 & 0.035 \\
    IP horizontal beta function & mm & \multicolumn{2}{c}{3.3} \\
    IP vertical beta function & mm & \multicolumn{2}{c}{0.1} \\
    IP horizontal beam size (rms) & nm & \multicolumn{2}{c}{729} \\
    IP vertical beam size (rms) & nm & \multicolumn{2}{c}{7.7} \\
    Average beam power delivered & MW & 8 & 2 \\
    Bunch separation & ns & \multicolumn{2}{c}{80} \\
    Average beam current & {\textmu}A & 16 & 64 \\[3pt]
    \hline
    & & \\[-15pt]
    \textit{RF linac parameters} & & & \\[-6pt]
    \hline
    & & \\[-15pt]
    Average gradient & MV/m & \multicolumn{2}{c}{25} \\
    Wall-plug-to-beam efficiency & \% & \multicolumn{2}{c}{50} \\
    RF power usage & MW & \multicolumn{2}{c}{47.5} \\
    Peak RF power per length & MW/m & \multicolumn{2}{c}{21.4} \\
    Cooling req.~per length & kW/m & \multicolumn{2}{c}{20} \\[3pt]
    \hline
    & & \\[-15pt]
    \multicolumn{2}{l}{\textit{PWFA linac and drive-beam parameters}} 
    & & \\[-6pt]
    \hline
    & & \\[-15pt]
    Number of stages &  & \multicolumn{2}{c}{16} \\[-8pt]
    Plasma density & cm$^{-3}$ & \multicolumn{2}{c}{$7\times10^{15}$} \\ %Plasma density & cm$^{-3}$ & \multicolumn{2}{c}{$1.5\times10^{16}$} \\
    In-plasma acceleration gradient & GV/m & \multicolumn{2}{c}{6.4} \\
    Average gradient (incl. optics) & GV/m & \multicolumn{2}{c}{1.2} \\
    Length per stage\tnote{a} & m & \multicolumn{2}{c}{5} \\
    Energy gain per stage\tnote{a} & GeV & \multicolumn{2}{c}{31.9} \\
    Initial injection energy & GeV & \multicolumn{2}{c}{5} \\
    Driver energy & GeV & \multicolumn{2}{c}{31.25} \\
    Driver bunch population & $10^{10}$ & \multicolumn{2}{c}{2.7} \\
    Driver bunch length (rms) & $\mu$m & \multicolumn{2}{c}{42} \\ %Driver bunch length (rms) & $\mu$m & \multicolumn{2}{c}{27.6} \\
    Driver average beam power & MW & \multicolumn{2}{c}{21.4} \\
    Driver bunch separation & ns & \multicolumn{2}{c}{5} \\
    Driver-to-wake efficiency & \% & \multicolumn{2}{c}{72} \\
    Wake-to-beam efficiency & \% & \multicolumn{2}{c}{53} \\
    Driver-to-beam efficiency & \% & \multicolumn{2}{c}{38} \\
    Wall-plug-to-beam efficiency & \% & \multicolumn{2}{c}{19} \\
    Cooling req.~per stage length & kW/m & \multicolumn{2}{c}{100} \\
    & & \\[-15pt]
    \hline
    & & \\[-17pt]
    \end{tabular}
%    \smallskip\footnotesize
\begin{tablenotes}
\item [a] The first stage is half the length and has half the energy gain of the other stages (see Section~\ref{sec:PWFAlinac}).
\end{tablenotes}
\end{threeparttable}
   \caption{Table of HALHF parameters.}
   \label{tab:2}
\end{table}
\renewcommand 
\thesubsection{\arabic{subsection}}
\renewcommand
\thesection{\Roman{section}.}
\subsection{Electron sources}
Sources of driver and colliding electron bunches are located at the start of the main RF and the plasma-based linacs, respectively. 
The beam-quality requirements are low for the drive beams and, because of the possibility for an emittance asymmetry (see Sec.~\ref{sec:lumiperX}), the colliding electron beam, and can be met by use of a photocathode injector. 
A dedicated electron damping ring is therefore not required~\cite{Xu_DR_Free}.
However, the colliding electrons should be as highly spin polarised as possible (the ILC electron gun was designed to produce 90\% polarisation), as this is highly advantageous for physics exploitation.
The photocathode sources are used to produce bunches of $1\times10^{10}$ electrons (1.6~nC) for collisions and
trains of drive-beam bunches with $2.7\times10^{10}$ electrons (4.3~nC) per bunch.

A train of 16 drive-beam bunches (one for each stage of the plasma-based linac) is produced for each colliding bunch, separated by 5~ns. 
The separation is determined by the rise times of kickers (2--4 ns) required to extract the beams and the transit time of the accelerated bunch between stages.
Many such trains are accelerated consecutively within a longer \textit{burst}.

\subsection{Positron source}
\label{sec:positronsource}
The HALHF uses a ``conventional" positron source~\cite{Chalkovska_positrons} such as that proposed for CLIC. 
The positrons produced by the source, which are unpolarised, are captured and transported to the conventional RF linac to be accelerated to 3~GeV before being transferred to the first and then second damping ring, which radiatively damp the emittance and energy spread to the required level.
In order to produce these positron bunches, dedicated electron bunches are produced and accelerated up to 5~GeV before being transferred via a return loop back to the positron target. 
Around $4\times10^{10}$ positrons (6.4~nC) are required per bunch.
A possible alternative scheme is to replace the dedicated 5~GeV electron beam with the spent 31.3~GeV positron beam after collisions (similar to the incident electron energy used at the Stanford Linear Collider), which if technically achievable, could lead to significant running-cost savings.

\subsection{Main RF linac for positron and drive-beam acceleration}
\label{sec:RFlinac}
The conventional linac, which is functionally split into two parts, serves two main purposes: to accelerate the electron drive bunches for the plasma-based linac to the required energy of 31.3~GeV; and to accelerate the positrons from 5~GeV to the final collision energy (also 31.3~GeV).

The gradient of this linac is assumed to be 25~MV/m, so that the total length required to accelerate to 31.3~GeV is 1.25~km.

This can, for instance, be achieved using an L-band normal-conducting RF linac. 
Another possibility would be to use a continuous-wave (CW) superconducting linac. For the purposes of this paper, we assume an L-band normal-conducting linac, with the bunch structure discussed below. 
If CW were preferable, the overall bunch structure would be different, i.e. continuous at 10 kHz, and would have consequences on the instantaneous heat load on the plasma accelerator. 
The total average power delivered to the drive beam train and positrons, if operated at 100~Hz (see Sec.~\ref{sec:bunchtrainpattern}), is 21.4~MW.

Operating at an energy efficiency estimated to be 50\%, and including the power for the positron source, requires approximately 48~MW of wall-plug RF power.
In the main linac, a peak power of 21.4~MW is required per metre of cavity during acceleration (see Section~\ref{sec:cost_running}), running for approximately 8--10~{\murm}s (one burst), which is compatible with conventional klystrons.

After acceleration, the electron driver train is separated from the positron bunch via a dipole and transferred to the plasma-based linac. 
Each positron bunch is accelerated with appropriate delay compared to the driver train, and at a phase offset of 180 degrees (similar to operation of the SLAC linac for the Stanford Linear Collider). 

\subsection{PWFA linac for high-energy electrons}
\label{sec:PWFAlinac}
The drivers travel via a turn-around loop to re-orientate their trajectory parallel to the colliding electron bunch, and are subsequently distributed by a set of kickers in a tree-like delay chicane \cite{PfingstnerIPAC2016} in such a way as to synchronize them with the passage of the colliding electron beam.

%--- FIGURE 2: plasma wakefield ---%
\begin{figure}[t]
	\centering\includegraphics[width=\linewidth]{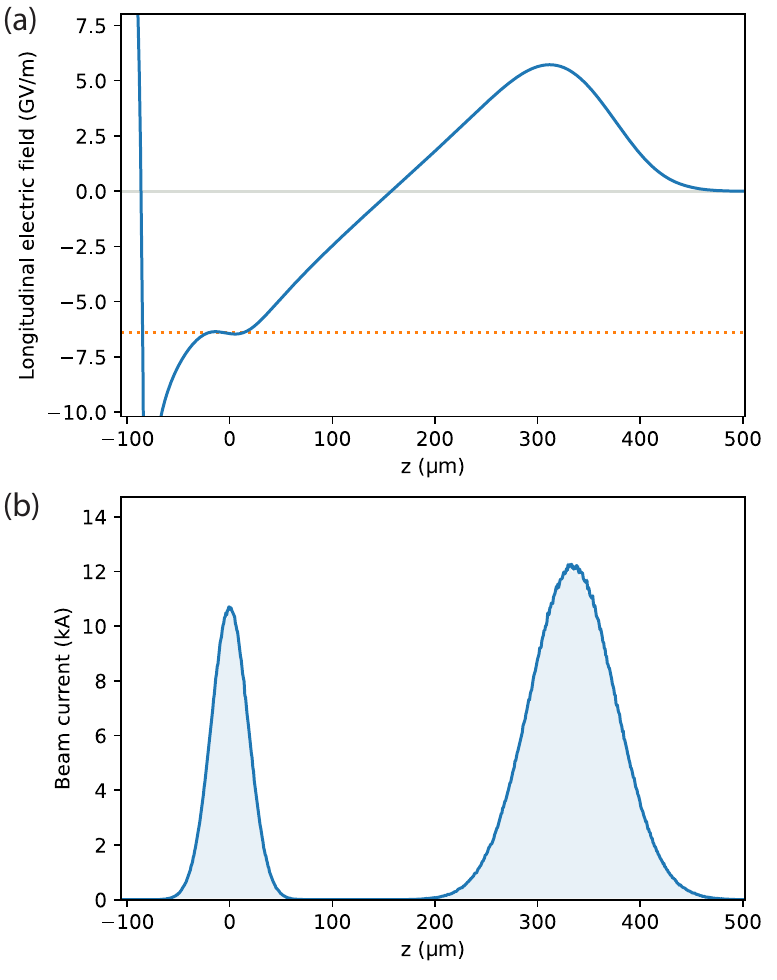}
	\caption{(a) Accelerating field in the proposed beam-driven plasma-accelerator stages, calculated using  Wake-T~\cite{FerranPousa2019WakeT}. Here, the plasma density is assumed to be $7\times10^{15}$~cm$^{-3}$. The orange dotted line corresponds to the nominal accelerating gradient of 6.4~GV/m. (b) Beam current of the electron driver (right) and trailing colliding bunch (left). The bunch lengths are 42 and 18~{\murm}m rms, respectively. In a 5~m long plasma-accelerator stage, the driver (31.3~GeV) loses 72\% of its total energy, 53\% of which is transferred to the colliding bunch via the plasma wakefield.}
    \label{fig:2_plasmawakefield}
\end{figure}

The first accelerating stage is short (half length) in order to ensure sufficient separation in energy between the colliding electron beam and the driver beam. 
The electron-source complex produces an electron beam with the required parameters and an energy of 5~GeV. 
The first plasma stage uses a 31.3~GeV driver to accelerate the electron beam to 21~GeV. The remaining stages are identical to each other; using 31.3~GeV drivers and assuming a conservative transformer ratio of about 1, the electron beam can be accelerated by just under 32~GeV per stage; consequently, 15 further stages are required to achieve 500~GeV beam energy.

The stages are separated by magnetic chicanes, used both to in- and out-couple the drivers, as well as to enforce a self-correction mechanism in longitudinal phase space \cite{Lindstrom_preprint_2021} for improved energy stability and damping of energy spread.
This self-correction mechanism also significantly improves the timing tolerances, such that state-of-the-art 10~fs~level driver synchronisation will probably be more than sufficient (detailed studies are required to determine the exact timing tolerance, which depends on staging optics).
Operating at a plasma density of around $7\times10^{15}$~cm$^{-3}$ and compressing the bunches to around 12~kA peak current gives an accelerating gradient of 6.4~GV/m (see Fig.~\ref{fig:2_plasmawakefield}).
The total length of plasma accelerator is only approximately 80~m (5~m per stage); the overall length of the plasma-based linac is dominated by the interstage optics (on average 26~m per stage, but scaling with the square root of the energy), resulting in a total length of approximately 410~m.
This gives an average accelerating gradient of 1.2~GV/m, which is more than an order of magnitude greater than in conventional RF accelerators.
A reduction in the accelerating gradient, provided the same energy gain can be delivered in each plasma cell (by extending its length) is therefore not a strong contributor to the total length of the facility.
Rather, the size of the facility is dominated by the high-energy beam-delivery system (see Section~\ref{sec:bds}).

The overall energy efficiency of each plasma-accelerator stage is estimated to be 38\%: 72\% of the energy of the driver goes into the wake, and 53\% of the wake energy is then extracted by the accelerated electrons.
These depletion and extraction efficiencies are not far beyond current experimental results \cite{LindstromPRL2021,PenaPRLsoon}, although they have not yet been shown simultaneously.
There is therefore a risk that the overall efficiency may be somewhat lower than assumed here.
Proceeding however on this assumption, a quarter of the power in the driver train is dumped into 16 separate beam dumps, corresponding to 5.3~MW total or 330~kW per beam dump\footnote{The first dump has to cope with more power than the others due to it being associated with a short cell.}.

Half of the remaining three quarters of the driver power is dumped into the plasma accelerators themselves, corresponding to 500~kW per stage or 100~kW/m.
This requires a significant, but probably manageable, cooling of the plasma cells. 
Cells that can cope with such energy depositions do not yet exist and must therefore be developed as a matter of urgency if the HALHF concept, or indeed any particle-physics-collider application with competitive luminosity, is to be realised.

Preserving the emittance from start to end will be one of the major challenges of implementing the plasma-based linac~\cite{Lindstrom_Thevenet}.
Fortunately, as discussed above, the possibility of an emittance asymmetry greatly increases the likelihood of success.
Assuming maximum asymmetry, normalised emittances can be as high as $160 \times 0.56$~mm-mrad.
To deliver this, mechanisms for suppressing transverse instabilities must be applied~\cite{LehePRL2017,Mehrling2018}.
In order to approximate the tolerances on transverse misalignment, we consider the last plasma-accelerator stage, where the accelerating electron beam is at its smallest transverse size (around $3 \times 0.2$ {\murm}m rms).
It therefore represents the worst-case scenario; assuming tens-of-percent emittance growth in this stage, we find that the misalignment tolerance will be of order 100~nm. 
This agrees with previous estimates of the required tolerance \cite{LindstromIPAC2016,Schulte2016RAST}; a tighter nominal constraint than for other designs (such as CLIC and ILC), but one that must be satisfied for a much shorter distance compared to conventional linacs. 
Moreover, and importantly, the combination of both charge and emittance asymmetry reduces the beam density  by $\sim$32 (2 in charge and 4 in each transverse plane), which largely mitigates the problem of ion motion \cite{RosenzweigPRL2005,AnPRL2017} within the colliding bunch~--~ions undergo only $\phi \approx 0.2$~radians of phase advance (which satisfies the requirement of $\phi \ll \pi/2$), assuming singly ionised argon ions.

The performance of the PWFA arm is clearly the least certain of all the elements in HALHF. If the accelerating gradient cannot be achieved, then more accelerating stages could in principle be added, which would require
a re-optimisation of the bunch pattern. If the beam quality of the PWFA beam were degraded compared with expectations, then the achievable luminosity would be reduced but to some extent could be recouped by increasing the number of colliding bunches (at the cost of increase power usage). The reduction of jitter to an acceptable level will be a major subject of the proposed R\&D. Given that the cost of the PWFA arm is essentially negligible in comparison with other parts of HALHF, and that the increases in performance requirements of the linac could not be extensive, there would be no sizeable effect on the overall cost of the project. 

Lastly, preserving the spin polarisation of the accelerating electrons is in principle possible in a plasma accelerator \cite{VieiraPRSTAB2011}, although this has not yet been experimentally verified.

%--- FIGURE 3: bunch train pattern ---%
\begin{figure}[t]
	\centering\includegraphics[width=\linewidth]{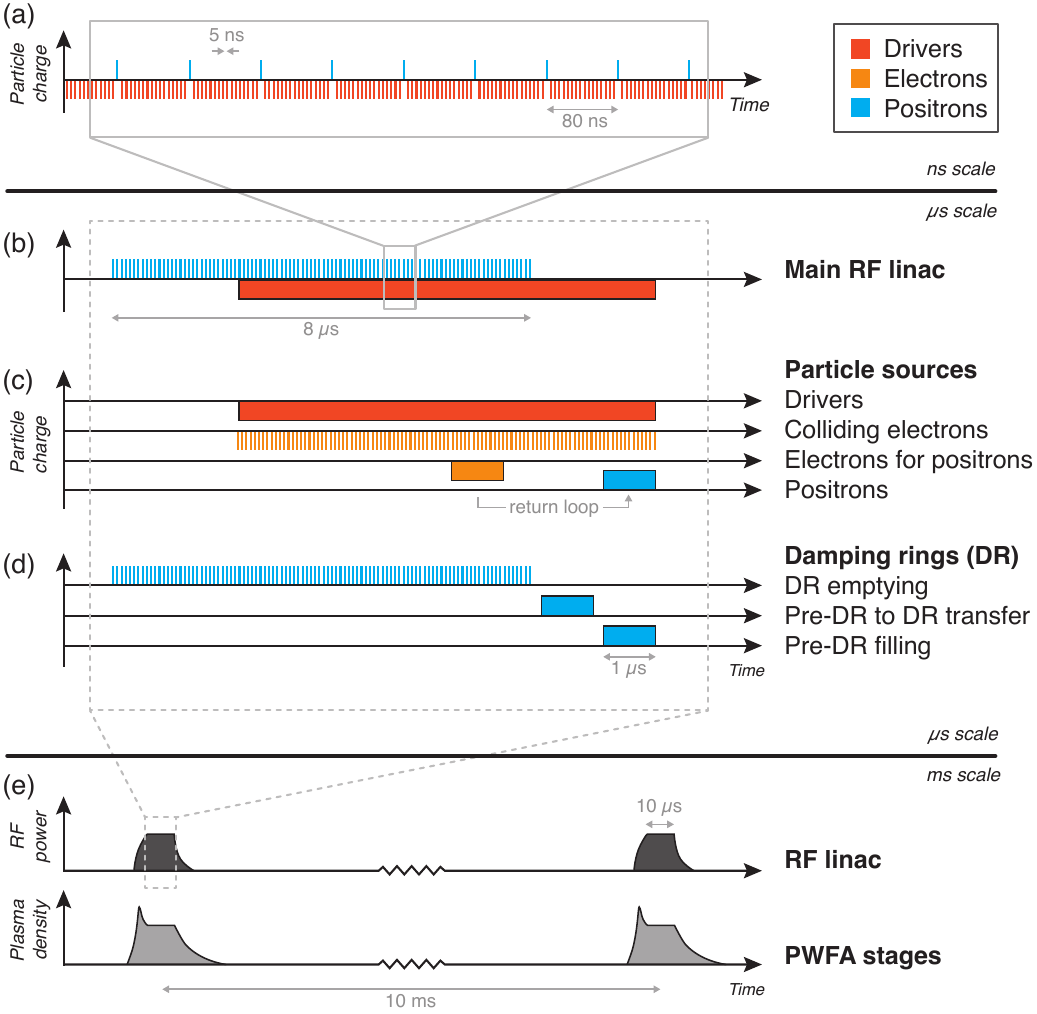}
	\caption{Bunch-train pattern, showing (a) the shortest relevant time structure (ns-scale) of the driver trains, interspersed with positron bunches, during acceleration in the the main RF linac, which (b) continues for the duration of a burst ({\murm}s-scale). The positron train must be shifted somewhat forward in time since it will traverse an additional turn-around loop. (c) Drivers, colliding electrons, and electron bunches delivered to the positron source are created at appropriate times. (d) The produced positrons are extracted from the damping ring at the appropriate separation, after which the pre-damping ring transfers its bunches to the damping ring, followed by the pre-damping ring being refilled. (e) The RF linac and PWFA stages operate with a $\sim$10~{\murm}s flattop at a repetition rate of 100~Hz (ms-scale).}
    \label{fig:3_bunchtrainpattern}
\end{figure}

\subsection{Overall bunch-train pattern}
\label{sec:bunchtrainpattern}
The bunch-train pattern is summarised in Fig.~\ref{fig:3_bunchtrainpattern}.
The number of drivers per colliding bunch is determined by the number of stages (i.e., 16).
The drivers are separated by approximately 5~ns (compared to 0.5~ns at CLIC), determined by the required delay between stages and the rise time of available kickers. 
This means that the shortest elapsed time between colliding bunches will be 80~ns, which is consistent with the maximum achievable repetition rate in a plasma accelerator~\cite{DArcyNature2022}.  

The length of the burst (i.e. a series of drive trains plus colliding bunch) is restricted by the duration over which a plasma accelerator can maintain constant acceleration~--~this is currently uncertain, but can be estimated to be on the 10~{\murm}s timescale \cite{Garland2021}. 

This restricts the number of colliding bunches per burst (separated by 80~ns) to around 100, with a resulting number of drivers per burst of 1600.

To compensate for the luminosity loss per bunch from beam--beam and hour-glass effects (Sec.~\ref{sec:lumiperX}), approximately 10,000 bunch collisions per second are required in order to achieve the same luminosity in the top 1\% of the energy spectrum as ILC (i.e., $0.4 \times 10^{34}$~cm$^{-2}$~s$^{-1}$).
This collision rate exceeds that for ILC (6,560~Hz), but is smaller than CLIC (15,600~Hz). 
Hence, the macro repetition rate of the bursts must be around 100~Hz, resulting in a requirement for the generation of approximately 160,000 drivers per second.
Similar to the CLIC two-beam scheme, the plasma-based linac works like a transformer, converting the high current (0.68~mA), low ``voltage" (31.3~GeV) drive beam to a low current (0.016~mA), high ``voltage" (500~GeV) colliding electron beam.

\subsection{Damping rings}
\label{sec:dampingrings}
In the HALHF scheme, there are two positron damping rings, but none for electrons.
The length of these damping rings is given by the product of the number of colliding bunches per burst and their required separation. 
We assume that the bunches must be separated by approximately 10~ns (ILC uses 8~ns \cite{ILC_TDR_Accelerator}), so that the 100 colliding bunches per burst will take up a total of 1~{\murm}s or 300~m. 
Additional gaps for kickers and superconducting wigglers must be added.
The CLIC main damping rings, on which our design is based, have a circumference of 359~m.
Our injection energy of 3~GeV is slightly larger than for CLIC (2.86~GeV), so we assume a radius of 400~m.
These damping rings have a characteristic damping time of approximately 1--2~ms~--~short enough to fully damp within the 10~ms period between bursts.
 
The particular filling and emptying pattern depends on the ring (see Fig.~\ref{fig:3_bunchtrainpattern}).
Since the time between collisions is 80~ns, the bunches must be kicked out of the main damping ring at this rate (i.e., every eighth bunch is extracted).
The pre-damping ring, however, can be filled and emptied continuously with bunches separated by 8~ns.
Moreover, the pre-damping ring must transfer its bunches to the main damping ring in the time between (or simultaneously with) the latter being emptied and the former filled.

It should be noted that there are considerable differences between the ILC and CLIC damping-ring parameters and those required for HALHF. 
For example, the number of bunches stored and therefore the size of the damping rings is considerably smaller than for ILC, whereas the positron bunch population is double; $4\times10^{10}$ as opposed to $2\times10^{10}$ particles for ILC.
The bunch population is also considerably larger than for CLIC (which has $5\times10^9$ particles), although the number of bunches and hence the damping-ring size is similar. 
The optimisation of damping-ring parameters is a complex process that we have certainly not attempted here. 
Our assumption is that such an optimised system, in terms of scope and cost, would not be very different from what is presented in this section.

\subsection{Beam-delivery systems}
\label{sec:bds}
The HALHF beam-delivery system (BDS) is, unsurprisingly, highly asymmetrical. 
It is modelled here on that for the ILC, which is designed for the same energy (500~GeV). 
A detailed design is beyond the scope of this paper; it is assumed that modifications to the general principles of the ILC design can be made at a later stage to produce the beta functions in Table~\ref{tab:1}.

Clearly, the HALHF positron BDS will be much shorter than that of ILC.
According to the prescription of Raimondi and Seryi~\cite{Raimondi_PRL_2001}, the final-focus part of the BDS length scales with energy somewhere between $E^{2/5}$ and $E^{7/10}$, depending on assumptions as to how the normalised emittance changes with energy. Other parts of the BDS scale proportional to energy or even the square of the energy~\cite{White_BDS}.
Each arm of the ILC BDS, which was designed to be upgraded to 500~GeV beams without changes in length, is around 2.25~km long.
Applying the scaling mentioned above implies that the HALHF positron BDS would be 320--740~m.
We emphasise that scaling by such a large factor is unreliable and a dedicated design should be carried out, but also that the effects of the uncertainty in the scaling on the overall cost and footprint of HALHF are negligible. For definiteness, we assume thatthe ILC lengths are scaled by $\sqrt E$, resulting in 
a positron BDS of 0.56~km.
Since the positrons must be turned around to collide with the electrons at the interaction point, a turn-around loop is required, which can in large part be combined with the positron BDS.
Finally, we note that the electron BDS must stretch the electron bunch from the short bunch length used in the PWFA stages (18~{\murm}m rms) to that required at the interaction point (75~{\murm}m rms). This decompression can be carried out within the long magnetic chicane used for energy-collimation in the BDS. From these estimates, it can be seen that the electron BDS is in fact the driver for the total length of the HALHF facility.
This strongly motivates further research into more compact beam-delivery systems.

\renewcommand
\thesection{\Roman{section}}
\section{Experimentation at HALHF}
\label{sec:Exp}
There has been very substantial design activity for detectors at a symmetric electron--positron Higgs factory~\cite{ILC_TDR_Detectors}. The basic design of such detectors can be taken over to HALHF with modifications to account for the boosted c.m. At HERA, the barrel and rear (i.e., lepton direction) detectors were essentially as would have been designed for a symmetric accelerator. In the forward (i.e., proton) direction, additional instrumentation was included, such as toroidal-field muon detection, greater depth of calorimetry and eventually, after detector upgrades, silicon disks placed perpendicular to the beam-pipe to aid the silicon barrel detectors placed around the interaction region~\cite{ZEUS_detector, H1_detector}. One beneficial aspect of an asymmetric collider is that the average decay distance of heavy quarks and leptons is also boosted, increasing tagging efficiencies. 

Generally speaking, none of these additions for an HALHF detector are either particularly costly or technically challenging and represent minor perturbations on existing design ideas. Indeed, it could be argued that since the final state of predominant interest is $HZ$, both of which are heavy, their decay products will be quite isotropic even at HALHF. 

In terms of electronics and triggering, the different bunch structure of HALHF could have implications. The overall triggering scheme should not change compared to that envisaged for the  ILC detectors, i.e.``triggerless'' running, recording all beam crossings. The viability of ``power-pulsing" techniques for silicon detectors at ILC~\cite{ILC-pp} would need to be re-evaluated, since there is no longer the very large gap between pulse trains characteristic of ILC. However, the HALHF bunch pattern is rather similar to that proposed for CLIC, where modified power-pulsing schemes have been proposed~\cite{CLIC-pp}. 

The one area that is substantially affected at the HALHF is the measurement of luminosity, vital for the calculation of any cross sections. At HERA, this was achieved by the measurement of the $ep \rightarrow ep\gamma$ bremsstrahlung process, known as Bethe--Heitler scattering \cite{Bethe_RoySoc_1934}. This is a QED process with a large and accurately calculable cross section in which both electron and photon are emitted at small angles to the incident electron direction. For electron--positron annihilation, luminosity is measured via the $e^+e^- \rightarrow e^+e^-$ elastic scattering process, known as Bhabha scattering. Again the cross section is strongly peaked in the direction of the incident particles. In a symmetrical collider, small calorimeters are placed as close to the beam pipe as practicable in order to maximise the rate. Coincidences between these detectors  eventually provide the precision determination of luminosity used in physics analyses. Often the single-hit rates are used to guide  accelerator operations in maximising luminosity.

For HALHF, the situation is slightly complicated by the boost along the electron direction. A similar situation was encountered in the experiments at B-factories such as BaBar and Belle, where the related process $e^+e^- \rightarrow \gamma \gamma$ is often also used. At Belle II, the cross section and luminosity are sufficiently large (as are the backgrounds close to the beam) that a reasonable measurement can be made with calorimeters that form part of the main detector, well away from the beam line~\cite{Bellelumi, BelleIIlumi}. However, for the B-factories, the boost is much smaller than at HALHF, e.g.~at Belle II\footnote{In fact, because the B-factories generally operate at the $\Upsilon(4s)$ resonance, which is produced approximately at rest, it is $\beta\gamma$ that is the operative quantity.}, $\gamma = 1.04$. In addition the cross section goes down as $s^{-1}$, so that measuring the luminosity in the central detectors is unlikely to give sufficient statistical precision. The provision of a dedicated small-angle detector in the electron direction is therefore necessary. Compared to a symmetric collider, the small-angle Bhabha cross section decreases by a factor $\sim (\theta \gamma)^{-2}$, where $\theta$ is the scattering angle. Even so, a dedicated detector close to the beam line behind a thin exit window should give sufficient rate both for online luminosity tuning and an accurate offline luminosity determination. The latter will be considerably strengthened by a coincidence with the scattered positron, which because of the boost, will predominantly be detected in the barrel. This also has the advantage that the energy and angle will be much more precisely measured than is the case in the positron-arm luminosity monitor of a symmetrical collider. 

\section{Cost Estimate}
\label{sec:cost}
\renewcommand
\thesection{\Roman{section}.}
The capital cost of the HALHF accelerator is estimated by appropriate scaling of sub-system costings given by other mature projects, principally ILC and CLIC, which have gone through extensive and detailed expert cost-review procedures. 
The running costs are dominated by wall-plug-power usage, which is estimated mostly by analogy to other projects.

\subsection{Capital cost}
\label{sec:cost_capital}
 The cost of HALHF is dominated by conventional subsystems and civil construction; the novel PWFA arm is relatively inexpensive, although delivering the requisite power is not. 
 Table~\ref{tab:3} summarises the capital cost of HALHF, broken down by subsystem and then scaled to analogous subsystems in other projects as indicated. The cost is converted as necessary into ILC units (ILCU)\footnote{1 ILCU is defined as \$1 on Jan.~1st, 2012.}. 
 Some explanatory remarks as to how the costing was carried out are given as footnotes to the table. 
Note that the HALHF costings scaled from ILC subsystems thereby include items such as vacuum systems, instrumentation, computing and other common services.
No attempt to cost the personnel requirement to build HALHF was made: suffice it to say that it should be considerably smaller than ILC.

\begin{table*}
\setlength\tabcolsep{0pt}
\centering
\begin{tabularx}{\textwidth}{|l|@{}>{\centering}X@{}|l|@{}>{\centering}X@{}|@{}>{\centering}X@{}|c|}
    \hline
    \rowcolor{gray!10} \hspace{3pt}Subsystem & Original cost (MILCU) & \hspace{3pt}Comment & Scaling factor & HALHF cost (MILCU) & \hspace{3pt}Fraction\hspace{3pt} \\
    \hline
    \hline
    \hspace{3pt}Particle sources, damping rings\hspace{3pt} & 430 & \hspace{3pt}CLIC cost~\cite{Charles2018}, halved for $e^+$ damping rings only\footnote{Swiss deflator from 2018 $\rightarrow$ 2012 is approximately 1. Conversion uses Jan 1st 2012 CHF to \$ exchange rate of 0.978.}\hspace{3pt} & 0.5 & 215 & 14\% \\
    \hline 
    \hspace{3pt}RF linac with klystrons & 548 & \hspace{3pt}CLIC cost, as RF power is similar & 1 & 548 & 35\% \\
    \hline
    \hspace{3pt}PWFA linac & 477 & \hspace{3pt}ILC cost~\cite{ILC_TDR_Accelerator}, scaled by length and multiplied by 6\footnote{Cost of PWFA linac similar to ILC standard instrumented beam lines plus short plasma cells \& gas systems plus kickers/chicanes. The factor 6 is a rough estimate of extra complexity involved.}  & 0.1 & 48 & 3\% \\
    \hline
    \hspace{3pt}Transfer lines & 477 & \hspace{3pt}ILC cost, scaled to the $\sim$4.6~km required\footnote{The positron  transfer line, which is the full length of the electron BDS, dominates; this plus two turn-arounds, the electron transport to the positron source plus small additional beam lines are costed.} & 0.15 & 72 & 5\% \\
    \hline
    \hspace{3pt}Electron BDS & 91 & \hspace{3pt}ILC cost, also at 500~GeV & 1 & 91 & 6\% \\
    \hline
    \hspace{3pt}Positron BDS & 91 & \hspace{3pt}ILC cost, scaled by length\footnote{The HALHF length is scaled by $\sqrt E$ and the cost assumed to scale with this length.} & 0.25 & 23 & 1\%  \\
    \hline
    \hspace{3pt}Beam dumps & 67 & \hspace{3pt}ILC cost (similar beam power) + drive-beam dumps\footnote{Length of excavation and beam line taken from European XFEL dump.} & 1 & 80 & 5\% \\
    \hline
    \hspace{3pt}Civil engineering & 2,055 & \hspace{3pt}ILC cost, scaled to the $\sim$10~km of tunnel required\hspace{3pt} & 0.21 & 476 & 31\% \\
    \hline
    
    \multicolumn{3}{c|}{} & \cellcolor{gray!10} Total & \cellcolor{gray!10} 1,553 & \cellcolor{gray!10} 100\% \\
    \cline{4-6}
\end{tabularx}
\caption{Estimated capital construction cost of the HALHF collider, broken down by subsystem. The costing is based on an appropriate scaling of the estimated costs of the equivalent CLIC, ILC or European XFEL subsystem. The total of 1.553 billion ILCU is equivalent to $\sim$\$1.9 billion today.}
\label{tab:3}
\end{table*}

Table~\ref{tab:3} shows that the total cost for HALHF is $\sim$1.55 billion ILCU.
We freely admit that this costing is very approximate, probably not better than 25\%. 
Nevertheless, we are convinced that it is substantially smaller than Higgs factories based on other technologies.
As for the cost of the experimental programme, we would expect the detector costs to be very similar to  those envisaged for either CLIC or ILC.
The extra instrumentation to cope with the boost in the forward direction should be minor.

The ILC accelerator costing is a decade old; ILCUs are interesting for comparison with ILC but do not represent the cost of a machine built today. Recently, the ILC has produced an updated costing in ILCUs for the Snowmass process in the USA~\cite{ILC_Snowmass}. This is based on a variety of improvements in the design. Some of these, such as reductions in the civil construction cost, would carry over directly to the HALHF cost estimate. Others, such as an increase in the baseline cavity acceleration gradient, would not. Even the smallest cost estimate for ILC remains much larger than that for HALHF.

A crude estimate of the HALHF cost "today" can be obtained by simply using the GDP deflator for the USA to update ILCUs into \$ of 2022 using a factor of 1.25. The capital cost of the HALHF collider ``today" would then be approximately \$1.9B.

The Implementation Task Force (ITF) report prepared for the Snowmass process~\cite{Snowmass_Implementation} examines details of many proposed collider projects.
In particular, it presents costings that are evaluated using a careful and sophisticated parameterisation process derived from the known costs of successful past projects and information from current component costs.
The ITF quotes the Total Project Cost (TPC) as required by the US Department of Energy, sometimes known as ``US accounting".
The costs given above for HALHF are in ``European accounting", where personnel costs, escalation etc.~are dealt with separately.
Although the ITF report does give figures for several PWFA-based concepts, none are useful for comparison with HALHF as they are for much higher energy.
However, the ITF TPC for an ILC Higgs factory, which is within the range \$7--12B, can be scaled in the same way as was done to estimate the capital cost of HALHF.
This gives a TPC for HALHF of \$2.3--3.9B in 2021 dollars. 
A recent explicit costing using the ITF methodology gives an estimate of \$4.46B~\cite{Spencer_private}.

\subsection{Running costs}
\label{sec:cost_running}
The HALHF running costs are dominated by the power used to produce the drive beams. 
The power required to produce and maintain the plasma is negligible. 
Accelerating 100 trains of 16 electron drivers (one for each plasma stage, see 
Section~\ref{sec:PWFAlinac}), each of which has 4.3~nC of charge, plus the positron bunches with 6.4~nC,
operated at a repetition rate of 100~Hz and 50\% wall-plug efficiency, requires around 48~MW of total wall-plug power.
Damping rings, of which there are two, add about 10~MW each~\cite{Charles2018}.

In addition to the high-level RF power, substantial cooling power is required, particularly for the PWFA linac. 
Without any detailed design for PWFA cells that can deal with the remnant power unavoidably deposited in the plasma, we assume that the system is similar to that of CLIC, which also drives one beam with another, although with very different technology. 
Excluding RF and magnets, the CLIC power budget is dominated by cooling, which adds roughly 50\% of the RF power requirement to the total. 
We assume a similar fraction for HALHF. 
On this estimate, the cooling requirement per meter of RF structure is approximately 20~kW/m, which is similar to that of the CLIC drive-beam linac.

The power requirement for HALHF from the sources mentioned above would therefore be $\sim$92~MW. 
Making a guess for magnet power, which will be substantially less than for CLIC, we round this up to 100~MW, roughly similar to ILC and CLIC Higgs Factories.

\renewcommand
\thesection{\Roman{section}}
\section{Possible staging and upgrade schemes}

Clearly any accelerator of HALHF's complexity needs to have a sizeable prototype. This needs to concentrate on the technologically advanced part, the PWFA linac. A scaled-down version of a few cells would first be constructed. 
This could immediately be applied in experiments in strong-field quantum electrodynamics (SFQED)~\cite{PoderPRX2018,ColePRX2018,DESY_workshop_2019,LUXECDR2021}, for which a multi-100~GeV electron beam would be very useful.
Once the appropriate parameters in the PWFA linac have been demonstrated, the remaining infrastructure could be constructed, in particular the positron and drive-beam sources, turn-arounds and BDS.
Throughout this upgrade, the PWFA linac could continue to operate SFQED experiments and then be connected to the full complex at the last possible moment.

Once the HALHF complex is complete, collisions at the $Z$ with the same $\gamma$ factor as for Higgs running could begin. 
This would allow accelerator and, in due course, detector commissioning in advance of full-power running. 

The HALHF concept, while excellent for a Higgs factory, is not competitive in luminosity with a dedicated circular $Z$-factory, which, with a relatively small radius optimised for $Z$ production would probably be of similar cost.
Neither is it optimal for very high energies.
As discussed above, the inefficiency in acceleration power increases with the boost $\gamma$.
If the positron energy is unchanged, the boost in the forward direction rapidly makes experimentation very difficult.
Nevertheless, reaching the top threshold, $\sim$350~GeV, while leaving the positron linac unchanged would require an electron energy of $\sim$980~GeV and a $\gamma$ of only $\sim$2.9. 
A more radical upgrade would be to leave the boost approximately unchanged at that of the Higgs factory by upgrading the energies of both the positron and PWFA linacs, to 44~GeV and 700~GeV, respectively. 
However, this would require radical changes to the site as the electron BDS, which dominates the accelerator footprint, would have to be increased in length. 
This would have significant knock-on effects for other parts of the complex. 
Of course, it would be possible to foresee this upgrade from the beginning and increase the various tunnel lengths to accommodate such an energy upgrade without requiring further civil construction. 
Such an upgrade would still be considerably cheaper than either an upgraded ILC or CLIC. Even though the advantages of the HALHF concept reduce as the CoM energy increases, it is still likely to be much smaller and cheaper than competing projects.

Another possibility is the construction of a $\gamma$--$\gamma$ collider~\cite{Higgs-factory-review}.
The positron source could be switched out and another electron PWFA linac and appropriate BDS constructed adjacent to it.  

A further upgrade in capability would be to replace the conventional positron source with an ILC-like scheme, using the spent electron beam to generate polarised positrons via a wiggler~\cite{ILC_TDR_Accelerator}. Schemes to generate high positron polarisation from electron beams of up to 500 GeV have been proposed~\cite{Mikhailichenko_2015}. Whether these can be adapted to the characteristics of the spent HALHF electron beam requires further consideration.

%% CONCLUSIONS
\section{Summary \& Conclusions}
\label{sec:Conclusions}

\begin{table*}[!hbtp]
    \begin{tabular}{p{0.25\linewidth}>{\centering}p{0.13\linewidth}>{\centering}p{0.13\linewidth}>{\centering\arraybackslash}p{0.13\linewidth}>{\centering\arraybackslash}p{0.13\linewidth}>{\centering\arraybackslash}p{0.13\linewidth}}
    \textit{Parameter} & \textit{Unit} & \multicolumn{2}{c}{\textit{HALHF}} & \textit{ILC} & \textit{CLIC} \\
    \hline
     & & $e^-$ & $e^+$ & $e^-$/$e^+$ & $e^-$/$e^+$ \\
    \hline
    Center-of-mass energy & GeV & \multicolumn{2}{c}{250} & 250 & 380 \\
    Center-of-mass boost &  & \multicolumn{2}{c}{2.13} & - & - \\
    Bunches per train & & \multicolumn{2}{c}{100} & 1312 & 352 \\
    Train repetition rate & Hz & \multicolumn{2}{c}{100} & 5 & 50 \\
    Average collision rate & kHz & \multicolumn{2}{c}{10} & 6.6 & 17.6 \\
    Average linac gradient & MV/m & 1200 & 25 & 16.9 & 51.7 \\
    Main linac length & km & 0.41 & 1.25 & 7.4 & 3.5 \\
    Beam energy & GeV & 500 & 31.25 & 125 & 190 \\
    Bunch population & $10^{10}$ & 1 & 4 & 2 & 0.52 \\
    Average beam current & {\textmu}A & 16 & 64 & 21 & 15 \\
    Horizontal emittance (norm.) & {\textmu}m & 160 & 10 & 5 & 0.9 \\
    Vertical emittance (norm.) & {\textmu}m & 0.56 & 0.035 & 0.035 & 0.02 \\
    IP horizontal beta function & mm & \multicolumn{2}{c}{3.3} & 13 & 9.2 \\
    IP vertical beta function & mm & \multicolumn{2}{c}{0.1} & 0.41 & 0.16 \\
    Bunch length & {\textmu}m & \multicolumn{2}{c}{75} & 300 & 70 \\
    Luminosity & cm$^{-2}$~s$^{-1}$ & \multicolumn{2}{c}{$0.81\times10^{34}$} & $1.35\times10^{34}$ & $2.3\times10^{34}$ \\
    Luminosity fraction in top 1\% & & \multicolumn{2}{c}{57\%} & 73\% & 57\% \\
    Estimated total power usage & MW & \multicolumn{2}{c}{100} & 111 & 168 \\
    Site length & km & \multicolumn{2}{c}{3.3} & 20.5 & 11.4 \\
    \hline
    \end{tabular}
   \caption{Comparison of main parameters between HALHF, ILC and CLIC, based on the latest updated values for each machine \cite{ILC_Snowmass,CLIC_Snowmass}. Note that the center-of-mass energy is 250~GeV for HALHF and ILC, but 380~GeV for CLIC.}
   \label{tab:4}
\end{table*}

We have sketched a concept for an electron--positron Higgs factory that avoids the difficulties inherent in the acceleration of positrons in plasmas, while producing a luminosity comparable to that given in the ILC TDR and the CLIC CDR ($0.81 \times 10^{34}$~cm$^{-2}$~s$^{-1}$ and $0.46 \times 10^{34}$~cm$^{-2}$~s$^{-1}$ in the peak 1\%) with a much reduced facility footprint, capable of fitting inside the boundaries of many current laboratories.
The main parameters of HALHF are compared to those of ILC and CLIC in Table~\ref{tab:4}. The ILC parameters shown in this table are upgraded compared to the TDR. 

The reduction in the size of HALHF compared to conventional Higgs factories leads to  a greatly reduced cost. 
This is achieved by utilising the very high gradients attainable in plasma-wakefield accelerators, greatly reducing the cost of the high-energy electron linac, and by minimising the cost of the conventional RF linac used to accelerate positrons.
The resulting asymmetric energies boost the annihilation products in the electron direction.
This should not cause a problem experimentally, as testified by the successful physics programme carried out at HERA, whose boost was larger than is proposed for HALHF.
Moreover, the asymmetric energy also enables use of asymmetric emittances, which makes it easier to achieve the required beam quality in the plasma-based accelerators and removes the need for a dedicated electron damping ring~--~saving significant capital and running costs.
  
We are of course aware that HALHF as outlined in this paper cannot be built tomorrow: many unsolved problems remain. 
Some R\&D on “conventional” systems is required, e.g.~on positron sources, damping ring optimisation and the detailed design of the beam-delivery system. 
However, the major challenge is to produce plasma accelerators with the characteristics required for HALHF.
We believe that the HALHF concept should act as a spur to the improvement of specific plasma-acceleration techniques.
In particular, there must be progress toward the use of multiple stages, self-correction mechanisms, higher accelerated charge (by a factor $\sim$10), higher repetition rate (by a factor $\sim$1000), plasma-cell design required to cope with large power dissipation (by a factor $\sim$1000) and reduction of beam jitter to an acceptable level (by a factor $\sim$10--100).
However, unlike ILC and CLIC, the problems to be solved are technical, not political.
There is likely to be a very different psychological impact of the HALHF price on nations considering hosting compared to that of ILC or CLIC.
The capital cost of \$1.9B is not much greater than the European XFEL in Hamburg, which was substantially borne by the host country.
Using US accounting, the lower estimate of the HALHF TPC cost of \$2.3--3.9B at 2021 prices overlaps with that of the Electron-Ion Collider at Brookhaven National Laboratory, which is essentially a national project. 

Although the HALHF concept is not competitive with circular colliders at lower energies and becomes increasingly more difficult and expensive with energy, it seems that HALHF may well be in the ``Goldilocks Zone"; just right for the Higgs boson. 
Although HALHF requirements are currently well beyond state-of-the-art, they may nevertheless be possible to achieve after a decade of intensive R\&D.
If this R\&D were successful, because of its compact size, HALHF could be constructed before many of the Higgs-factory projects currently under consideration. The necessary R\&D should therefore be vigorously pursued as soon as possible. 
%If the required technical improvements can be achieved, this would be a triumph not only for the plasma-wakefield community, generating countless applications throughout science, but also for particle physics. 

%% ACKNOWLEDGEMENTS
\vspace{0.5cm}
\begin{acknowledgments}

We are grateful to M.~Berggren for running the initial GUINEA-PIG simulations.
We thank N.~Walker for many helpful conversations and him, E.~Adli, S.~Gessner, S.~Stapnes, D.~Schulte, M.~Wing and A.~Wolski for perceptive comments on the manuscript. 
BF is grateful to the Leverhulme Trust for the award of an Emeritus Fellowship and the University of Hamburg/DESY for financial support. 
This work was supported by the Research Council of Norway (NFR Grant No.~313770).

\end{acknowledgments}

Data availability statement: Any data that support the findings of this study are included within the article.

%% BIBLIOGRAPHY
\bibliographystyle{apsrev}

\end{document}